\documentclass[12pt,preprint]{aastex631}

\usepackage{natbib}
\usepackage{graphicx}
\usepackage{url}
\usepackage{indentfirst}
\usepackage{gensymb}
\usepackage{textcomp}
\usepackage{amsmath}
\usepackage{physics}
\usepackage{listings}
\usepackage[title]{appendix}

\makeatletter
\def\@to{to}
\makeatother

\DeclareGraphicsExtensions{.png,.jpg}
\usepackage[utf8]{inputenc}

\newcommand{\h}{$h_{70}^{-1}$}
\newcommand{\be}{\begin{equation}}
\newcommand{\ee}{\end{equation}}
\newcommand{\ba}{\begin{eqnarray}}
\newcommand{\ea}{\end{eqnarray}}

\newcommand{\simgt}{\lower 2pt \hbox{$\, \buildrel {\scriptstyle >}\over {\scriptstyle\sim}\,$}}
\newcommand{\simlt}{\lower 2pt \hbox{$\, \buildrel {\scriptstyle <}\over {\scriptstyle\sim}\,$}}

\newcommand{\ls}{\lower 2pt \hbox{$\;\scriptscriptstyle \buildrel<\over\sim\;$}}
\newcommand{\gs}{\lower 2pt \hbox{$\;\scriptscriptstyle \buildrel>\over\sim\;$}}

\def\apjl{\rmfamily{ApJ~}}        
\def\aap{\rmfamily{A\&A~}}        
\def\apjs{\rmfamily{ApJS~}}       

\def\nodata{ ~$\cdots$~ }

\begin{document}
\title{Multi-wavelength and Environmental Properties of Variability Selected Low Luminosity Active Galactic Nuclei}
\author[0000-0002-7720-3418]{Heechan Yuk}
\affil{Homer L.\ Dodge Department of Physics and Astronomy,
University of Oklahoma, Norman, OK 73019, USA}
\email{hyuk@ou.edu}

\author[0000-0001-9203-2808]{Xinyu Dai}
\affil{Homer L.\ Dodge Department of Physics and Astronomy,
University of Oklahoma, Norman, OK 73019, USA}
\email{xdai@ou.edu}

\author{Marko Mićić}
\affil{Homer L.\ Dodge Department of Physics and Astronomy,
University of Oklahoma, Norman, OK 73019, USA}

\begin{abstract}
We present the multi-wavelength and environmental properties of 37 variability-selected active galactic nuclei (AGNs), including 30 low luminosity AGNs (LLAGNs), using a high cadence time-domain survey (ASAS-SN) from a spectroscopic sample of 1218 nearby bright galaxies.  We find that high-cadence time-domain surveys uniquely select LLAGNs that do not necessarily satisfy other AGN selection methods, such as X-ray, mid-IR, or BPT methods. In our sample, 3\% of them pass the mid-infrared color based AGN selection, 18\% pass the X-ray luminosity based AGN selection, and 60\% pass the BPT selection. This result is supported by two other LLAGN samples from high-cadence time-domain surveys of TESS and PTF, suggesting that the variability selection method from well-sampled light curves can find AGNs that may not be discovered otherwise.
These AGNs can have moderate to small amplitudes of variability from the accretion disk, but, of many of them, with no strong corona, emission lines from the central engine, or accretion power to dominate the mid-IR emission. The X-ray spectra of a sub-sample of bright sources are consistent with a power law model.
Upon inspecting the environments of our sample, we find that LLAGNs are more common in denser environments of galaxy clusters in contrast with the trend established in the literature for luminous AGNs at low redshifts, which is broadly consistent with our analysis result for luminous AGNs limited by a smaller sample size. 
This contrast in environmental properties between LLAGN and luminous AGNs suggests that LLAGNs may have different trigger mechanisms.
\end{abstract}

\section{Introduction}
It is now believed that most galaxies have supermassive black holes (SMBHs) at their centers \citep[e.g.,][]{kormendy95, magorrian98}. Some of them are known to be actively accreting nearby matter, releasing a tremendous amount of energy during the process. These are known as active galactic nuclei (AGNs) and they are some of the most energetic phenomena in the observable universe. These are fascinating objects to study, as they provide unique opportunities to study accretion physics and galaxy evolution.

Many methods are used to identify AGNs, such as emission lines in optical spectra \citep[e.g.,][]{baldwin81, kewley06}, X-ray luminosity \citep[e.g.,][]{mushotzky04}, UV/optical colors \citep[e.g.,][]{schmidt83, richards02}, mid-infrared colors \citep[e.g.,][]{stern05, assef10, stern12}, and radio emission \citep[e.g.,][]{tadhunter16}. Among many methods, variability selection \citep[e.g.,][]{macleod11, baldassare20, yuk22} has been gaining popularity in recent years with the availability of time-domain surveys.

Variability across the wide range of electromagnetic spectrum and timescales is one of the defining features of AGNs. However, the mechanism behind is still a topic of debate. There exist many theories on the origin of AGN variability, including thermal processes \citep[e.g.,][]{lightman74, shakura76, kelly09}, accretion disk instabilities \citep[e.g.,][]{rees84, kawaguchi98, trevese02}, photon reprocessing \citep[e.g.,][]{haardt91, mchardy18}, and supernovae \citep[e.g.,][]{terlevich92, aretxaga97}. What is known, however, is that the AGN variability is stochastic \citep[e.g.,][]{kelly09, macleod10}. This key characteristic is used to determine whether a galaxy displays an AGN-like variability. For example, \citet{macleod11} used a damped random walk to model the AGN variability and analyzed the Sloan Digital Sky Survey (SDSS) Stripe 82 light curves of $\sim$10,000 variable objects to determine which ones display AGN-like variability. They were able to achieve the completeness of $>$90\%, showing the robustness of this method. 

Variability selection is becoming a more powerful tool as more high-quality photometry data is becoming available, enabling the discovery of more AGNs, especially in the low-luminosity regime.
Many early variability selection studies used the SDSS Stripe 82 light curves \citep[e.g.,][]{sesar07, schmidt10, butler11, macleod11}. On average, the light curves \citet{macleod11} used had over 60 epochs spanning over 10 years.
Studies that followed used larger light curve databases. \citet{sanchezsaez19} used QUEST-La Silla Survey light curves which had about 120 epochs over 3.5 years on average. \citet{baldassare20} used Palomar Transient Factory (PTF) light curves with about 65 epochs over 4 years on average. \citet{yuk22} used All-Sky Automated Survey for SuperNovae (ASAS-SN) light curves with about 280 epochs over 5.5 years in V-band and 180 epochs over 1.6 years in g-band on average. \citet{treiber23} used the high cadence Transiting Exoplanet Survey Satellite (TESS) light curves with more than 1000 epochs over 30 days.  With the better-sampled light curves, these new variability studies have selected AGNs with moderate to small variability amplitudes and are able to probe the low luminosity AGN (LLAGN) population.
Variability-based AGN selection in infrared (IR) light curves also demonstrates the evolution of this method. There are many studies on IR variability of AGNs \citep[e.g.,][]{neugebauer89, sanchez17, son22}. However, it was only recently that IR variability selection became a viable method to find AGNs, as surveys gathered sufficient data \citep[e.g.,][]{elmer20, green24}.

LLAGNs are an interesting subset of AGNs, as they are speculated to have different trigger mechanisms \citep[e.g.,][]{hopkins09, hopkins14} and accretion structure \citep[e.g.,][]{ho97, nicastro00, laor03, elitzur06}. 
LLAGNs are commonly selected using the BPT or X-ray methods \citep[e.g.,][]{kewley06, pellegrini07, ho08, wrobel08}, and variability adds an additional channel \citep[e.g.,][]{yuk22, messick23, treiber23, wasleske23}.

AGN population and evolution are also observed to have environmental dependence. Several studies report that the AGN fraction among galaxies in clusters is lower than that of among field galaxies at low redshifts \citep[e.g.,][]{kauffmann04, koulouridis10, lopes17, mishra20, koulouridis24}. Within a galaxy cluster, the AGN fraction is observed to increase with the distance from the cluster center \citep[e.g.,][]{koulouridis18, koulouridis24}. However, there are also other studies that observe no significant correlation between the local galaxy density and the AGN fraction \citep[e.g.,][]{miller03, silverman09, pimbblet13}. \citet{man19} find that the AGN fraction decreases with increasing overdensity of nearby galaxies, but is less dependent on the number of nearby star-forming galaxies.
At higher redshifts, the overall AGN fraction is observed to increase \citep[e.g.,][]{eastman07, haggard10, martini13, yang18}. \citet{eastman07} report that the AGN fraction in clusters increases more rapidly with redshift than the field AGN fraction.

In this study, we examine the multi-wavelength and environmental properties of the galaxy and variability selected AGN sample of \citet{yuk22}, where 70\% of their AGN candidates are LLAGNs. From this analysis, we seek to constrain the overlap fraction among different AGN selection methods, and AGN fractions among field and cluster galaxies, especially for LLAGNs.  
In Section \ref{secmethods}, we introduce the sample and the methods we use to analyze. In Section \ref{secresults}, we present the results. In Section \ref{secdiscussion}, we discuss the implications of the results. For this study, we assume the $\Lambda$CDM cosmological model with $\Omega_\Lambda=0.7$, $\Omega_M=0.3$, and $H_0=70$ km s$^{-1}$ Mpc$^{-1}$.

\section{Methods}
\label{secmethods}

\subsection{Sample}
\citet{yuk22} analyzed the All-Sky Automated Survey for SuperNovae (ASAS-SN) light curves of 1218 bright galaxies with Sloan Digital Sky Survey (SDSS) spectra. These galaxies are nearby, where 98\% of them have the redshift of less than 0.05. They used the excess variance and structure function to determine if a given light curve has an AGN-like variability. They found 37 of them to be variability-selected AGNs, 30 of which are LLAGNs. Upon analyzing the spectra, they found that about 10\%-30\% of the variability selected AGNs lie in the star-forming region in BPT diagrams, suggesting that the variability selection method can find AGNs that spectral classifications may miss. 
We note that the luminous AGN sample may be incomplete because the SDSS spectroscopic sample can exclude local luminous AGNs \citep{strauss02}. However, the comparisons of LLAGN or luminous AGN in different environments are not affected by this bias.
We further study the infrared, X-ray, and environmental properties of this AGN sample.

\subsection{Infrared classification}

\citet{assef10} analyzed the mid-infrared colors of galaxies and devised a set of criteria to select AGNs using colors from Wide-field Infrared Survey Explorer (WISE; \citealt{wright10}). Their criteria are:
\begin{equation}
    W1 - W2 > 0.85,
\end{equation}
\begin{equation}
    W3 - W4 > 2.1,
\end{equation}
and
\begin{equation}
    W1 - W2 > 1.67 (W3 - W4) - 3.41,
\end{equation}
where $W1$, $W2$, $W3$, and $W4$ are Vega magnitudes in the WISE wavelength bands at 3.4, 4.6, 12, and 22 microns, respectively. \citet{stern12} suggest a simpler criterion: $W1-W2\ge0.80$. We use both criteria to check our sample to compare the AGN selection methods.

\subsection{X-ray detection}

Many AGNs are known to be strong sources of X-ray emission. X-ray luminosity of 10$^{42}$ erg s$^{-1}$ is commonly used as the threshold to determine if a galaxy is an AGN \citep{mushotzky04}. To test this criterion and compare with the variability selection method, we gathered X-ray data from a multitude of sources.

One of the X-ray data sources is the extended ROentgen Survey with an Imaging Telescope Array (eROSITA) which was built for an all-sky survey. The German eROSITA Consortium released the data to the public for the western galactic hemisphere ($179.9423568<l<359.94423568$) on the eROSITA-DE Data Release 1 \citep{merloni24}. Out of the 1218 sample galaxies, 614 of them are within this hemisphere, 22 of which are the variability selected AGNs. We computed the X-ray luminosity based on the flux in energy range of 0.2--8 keV. For non-detections, we obtained the upper limits in 0.2--5 keV energy range \citep{tubinarenas24}.

In addition to eROSITA-DE, we searched for X-ray counterparts to our LLAGN in the Chandra archival database. Appropriate Chandra observations have been reduced and analyzed using the most recent version of the Chandra Interactive Analysis of Observations tool, CIAO 4.17, and calibration database CALDB 4.11.6. We started by downloading and reprocessing Chandra data, followed by cutting out small 60$\times$60 arcsecond square images centered on galaxies of interest. 
With the X-ray flux and redshift, we calculated the X-ray luminosity with K-correction and galactic absorption correction.

\subsection{AGN environment}

We examined the global and local environment of the sample to check for the AGN fraction's environmental dependence, where we refer to global as a region equivalent to a cluster or group size and local as the region containing the nearest neighbors of the source.

We used two methods to check the global environment.
First, we checked if the sample belongs to a galaxy cluster. For this, we used the catalog of galaxy clusters of \citet{wen24} (WH24). WH24 includes approximately 1.58 million galaxy clusters within the redshift of 1.5 based on the Dark Energy Spectroscopic Instrument (DESI) Legacy Imaging Surveys. \citet{wen24} also report the cluster redshifts, where about 330,000 of them are spectroscopic.
WH24 covers most of the sky except for the galactic plane.
At redshifts comparable to the sample ($z\lesssim0.10$), the number of galaxy clusters in WH24 is 4955, where 56\% of them having spectroscopic redshift measurements.

Because photometric redshifts typically have large uncertainties, it is difficult to precisely measure the physical distance between the WH24 clusters and the sample galaxies. Because of this, to avoid incompleteness, we use a 2-D matching method. For a given sample galaxy, we check WH24 clusters within a redshift difference of $\pm0.05$ of the target galaxy. Then we check the angular separation that corresponds to 2 Mpc given its distance. Using this method, we determined whether the galaxies in the sample are in clusters.
To check the robustness of the redshift range of $\pm0.05$, we also tested with $\pm0.03$.

Second, we checked the number of galaxies within 2 Mpc. We used NASA Sloan Atlas (NSA) for this, which is a galaxy of local galaxies ($z<0.15$) derived from SDSS data \citep{abdurrouf22}. Though this catalog does not cover all sky and all redshifts, because the sample is also local and from the SDSS spectroscopic sample, we do not suffer from significant incompleteness. Since the sample and the NSA galaxies have spectroscopic redshift measured, it is possible to calculate the distance from a sample galaxy to catalog galaxies by combining the angular separation and redshift offset. 

To examine the local environment, we computed two parameters using the NSA galaxies: distance to the nearest galaxy and the tidal index. The tidal index is a measurement of galaxy interactions \citep[e.g.,][]{karachentsev99, karachentsev04, pearson16}, essentially the density of stellar mass in a sphere to the nearby massive member. We use the version of the tidal index by \citet{pearson16},

\begin{equation}
    \Theta = \log\bigg(\frac{M_*/(10^{11}M_{\odot})}{(D/\textrm{Mpc})^3}\bigg),
\end{equation}
where $M_*$ is the stellar mass of the neighboring galaxy and $D$ is the distance to that galaxy. For each galaxy in the sample, we computed the tidal index with all NSA galaxies and took the maximum value.

\section{Results}
\label{secresults}

\subsection{Infrared classification}

\begin{figure}
    \centering
    \includegraphics[width=4.5in]{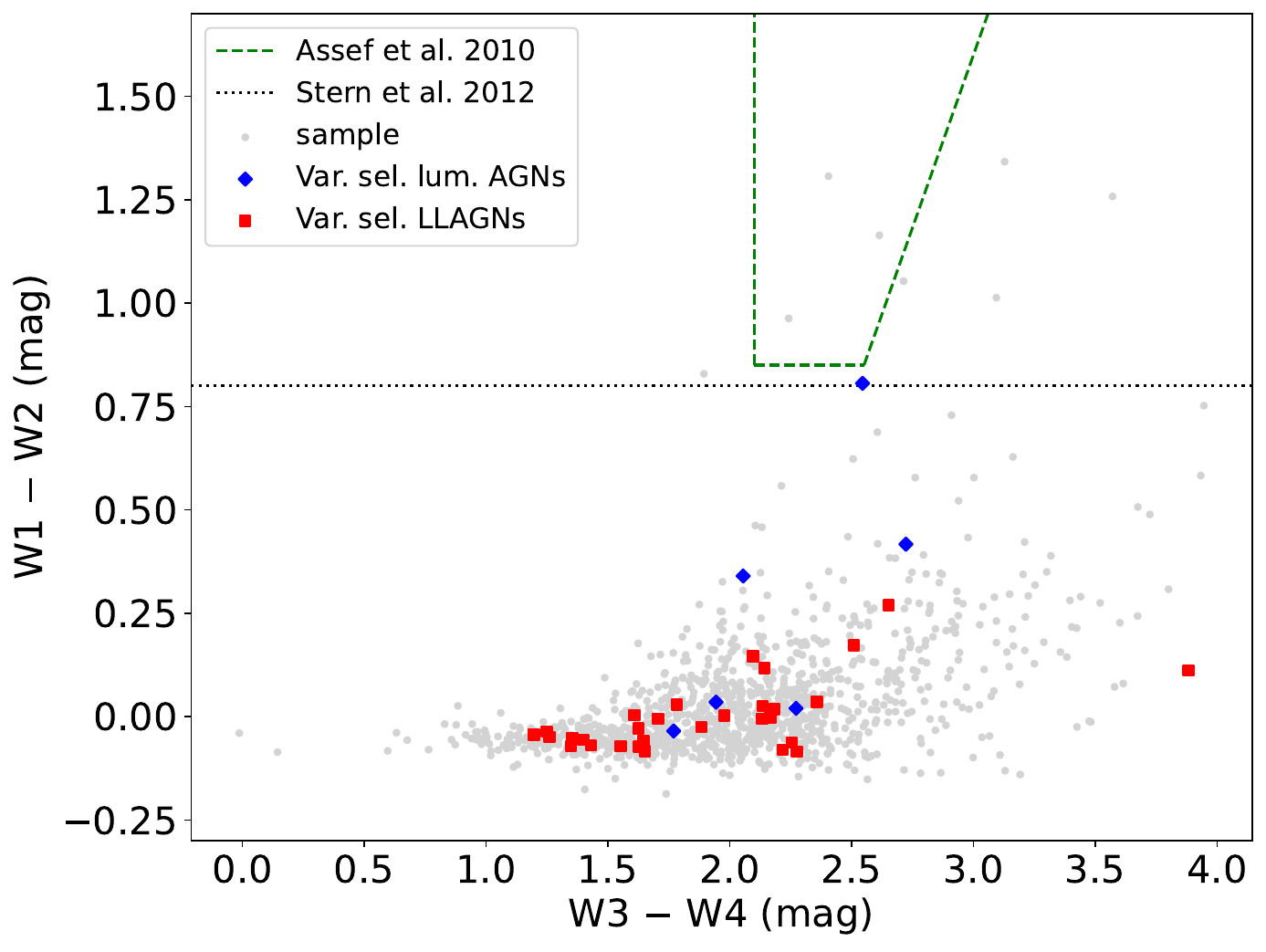}
    \caption{WISE color distributions of the variability selected AGNs and their parent galaxy sample with the IR color-based AGN selection boundaries labeled. Out of 36 variability selected AGNs with WISE measurements in all 4 wavelength bands, only 1 is classified as an AGN by the criterion of \citet{stern12}.}
    \label{wise}
\end{figure}

1187 out of 1218 total targets (36 out of 37 variability selected AGNs) have WISE data available.
When we check the infrared colors, we find that three lie within the AGN color criteria of \citet{assef10}, none of which are variability selected AGNs. If we use the broader criterion of \citet{stern12}, there are 9 targets that satisfy the condition, 1 of which is a variability selected AGN. In other words, only about 3\% of the variability selected AGNs are classified as IR AGNs. The infrared color-color diagram is shown in Figure \ref{wise}. We individually examined the ASAS-SN light curves of galaxies that are classified as AGNs by IR colors but not by variability selection and found that they do not have significant variability above the detection threshold to be included in the variability-selected AGN sample.

\subsection{X-ray detection}

\begin{figure}
    \centering
    \includegraphics[width=4.5in]{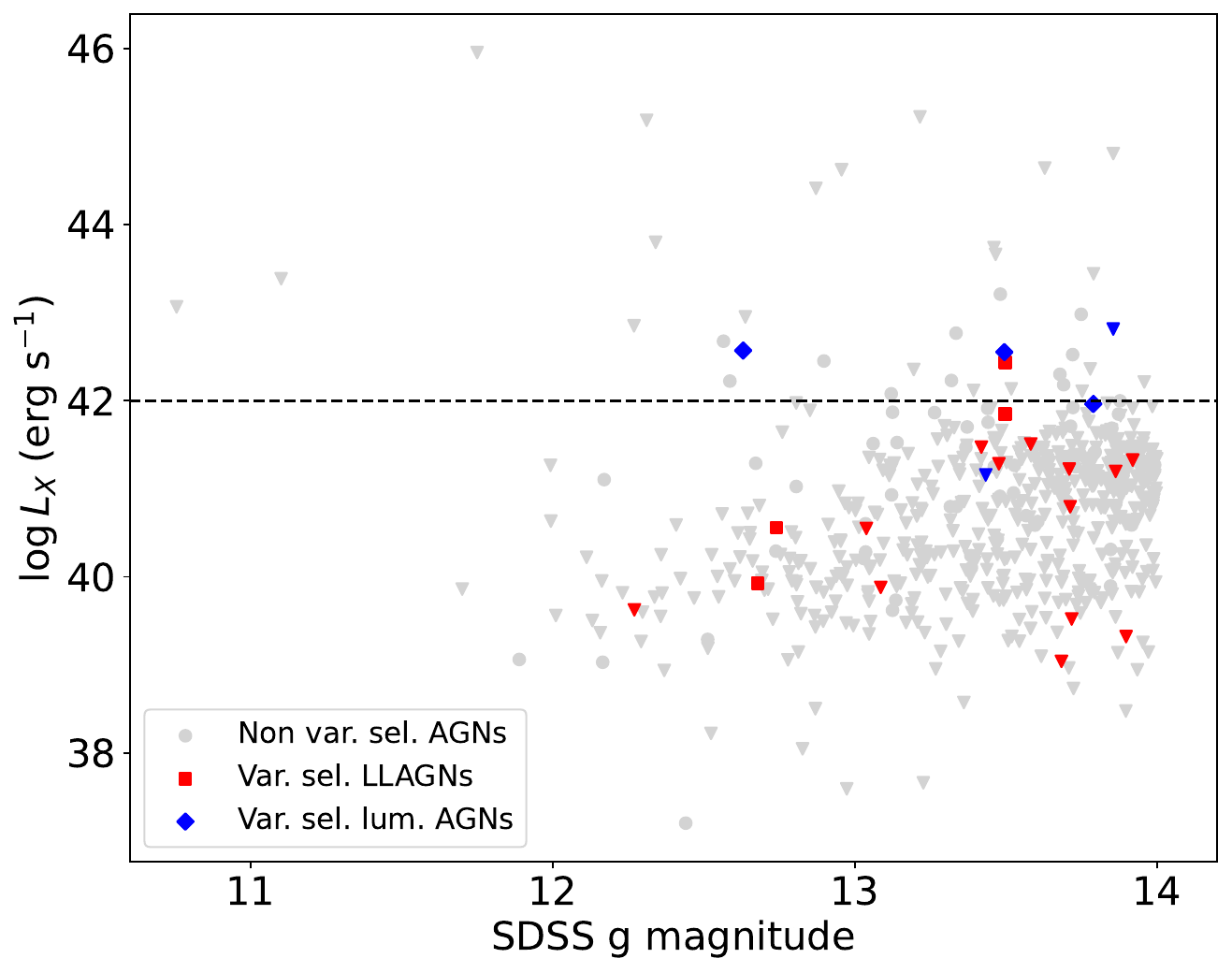}
    \caption{X-ray luminosity distribution of the variability selected AGNs and their parent galaxy sample based on eROSITA-DE data. The red squares, blue diamonds, and gray circles represent the X-ray detections of variability selected LLAGNs, luminous AGNs, and galaxies.  The upper limits are labeled with downward triangles. The dashed line indicates the luminosity limit for AGNs ($L_X>10^{42}$ erg s$^{-1}$). Out of 22 variability selected AGNs in the eROSITA-DE hemisphere, 7 had detections, and 3 are bright enough to be classified as AGNs.}
    \label{erositadist}
\end{figure}

We find that 76 of 614 targets in the western galactic hemisphere (7 out of 22 variability selected AGNs in that hemisphere) have eROSITA detections. The variability selected AGNs with eROISTA detections are NGC 4043, NGC 4061, NGC 4066, NGC 4224, NGC 4233, NGC 4235, and NGC 5162. 
16 sources, including 4 of the variability selected AGNs (NGC 4043, NGC 4061, NGC 4066, and NGC 4235) have the X-ray luminosity greater than $10^{42}$ erg s$^{-1}$.
When checking the upper limits, we find that 24 targets potentially have the X-ray luminosity of $>10^{42}$ erg s$^{-1}$, where 1 of them is a variability selected AGN.
From Chandra, we uncovered three nuclear point X-ray sources that are likely related to AGN activity associated with NGC 4224, NGC 4233, and NGC 5273, and their unabsorbed 0.2-10 keV luminosities are $3\times10^{40}$, $5\times10^{40}$, and $5\times10^{40}$ erg s$^{-1}$, respectively. 
The distribution of eROSITA X-ray luminosity is shown in Figure \ref{erositadist}.

\begin{figure}
    \centering
    \includegraphics[width=4.5in]{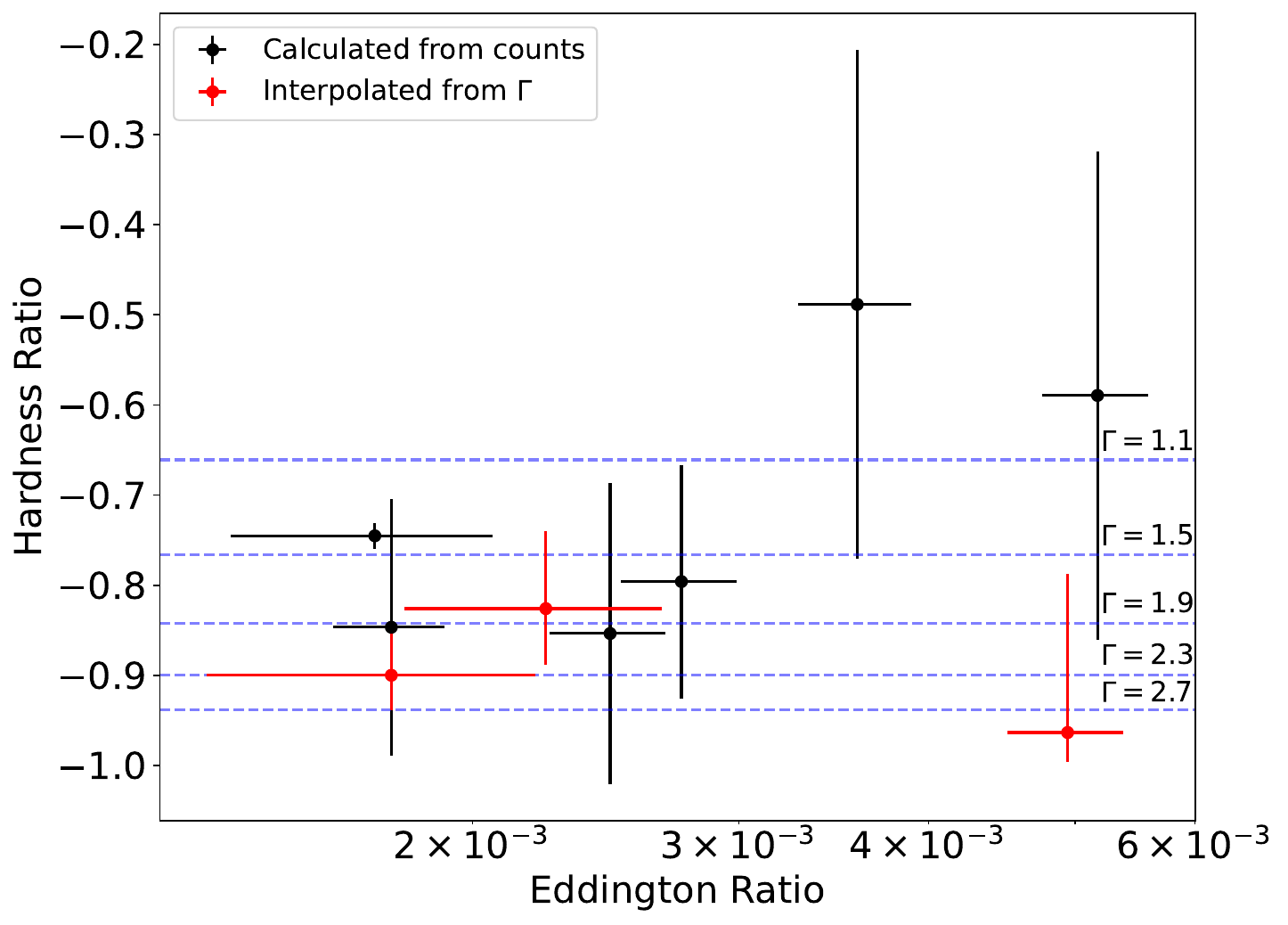}
    \caption{Hardness ratio versus Eddington ratio of X-ray detected variability selected AGNs. The blue dashed lines indicate the approximate hardness ratio corresponding to a set of photon indices by assuming a powerlaw model modified by Galactic absorption. Red points show the photon indices of three sources from the spectral fits of Chandra data, converted to equivalent hardness ratios.}
    \label{hardness_ratio}
\end{figure}

We also calculated additional X-ray properties of the variability selected AGNs. We computed the hardness ratio, 

\begin{equation}
    HR=\frac{C_H-C_S}{C_H+C_S},
\end{equation}
where $C_H$ and $C_S$ are photon counts in hard and soft bands, respectively, with the boundary at 2 keV. From the 7 variability selected AGNs with X-ray detections, one of them did not have sufficient counts to compute this ratio. We plotted hardness ratio against Eddington ratios in Figure \ref{hardness_ratio}.

\begin{figure}
    \centering
    \includegraphics[width=6.5in]{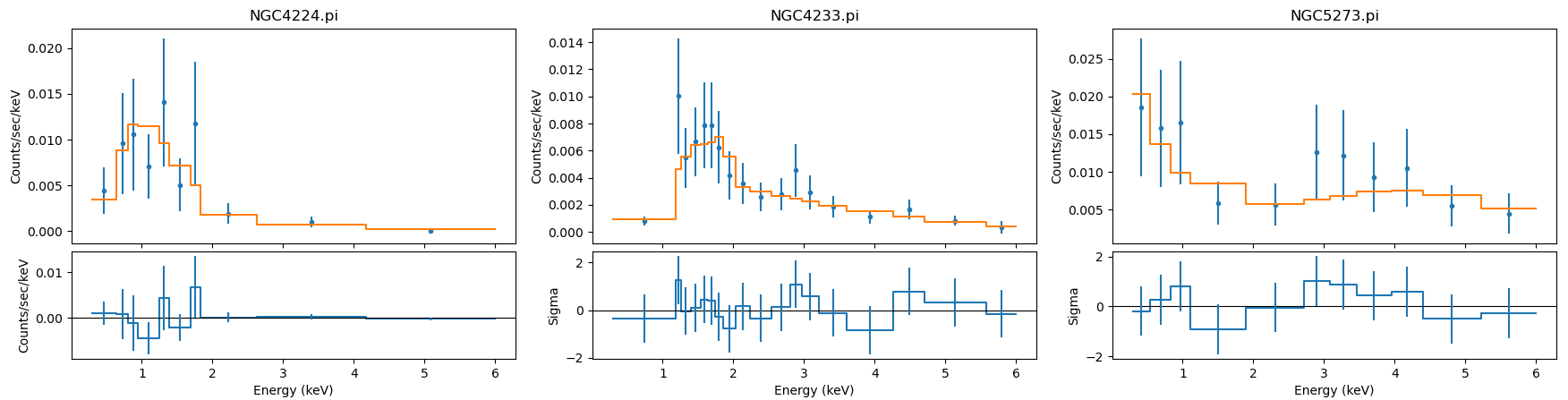}
    \caption{The Chandra spectral fits for NGC 4224, NGC 4233, and NGC 5273 with a power law model modified by Galactic absorption, where the photon index is constrained as $2.3\pm0.4$, $1.8\pm0.4$, and $3.1\pm1.5$, respectively. An additional blackbody component is needed to fit the NGC~5273 spectrum.  The bottom panels show the residuals.}
    \label{xspecfit}
\end{figure}

For Chandra sources, we performed spectral fitting analysis of the three sources using SHERPA, CIAO’s modeling and fitting application. NGC 4233 was fitted using the simple absorbed power law model, while NGC 5273 required an additional black body component. NGC 4224, the faintest source with only 28 net counts, was fitted using the simple absorbed power law model and Cash statistics. This approach is usually adopted for low-count X-ray spectral fitting \citep{marchesi16, mezcua18, micic22} since it does not require count binning in the spectral analysis. We find values of photon indices for NGC 4233, NGC 5273, and NGC 4224 to be $1.8\pm0.4$, $3.1\pm1.5$, and $2.3\pm0.4$. The spectral fits of these three targets are shown on Figure \ref{xspecfit}.
The summary of spectral fitting procedure and AGN X-ray properties is given in Table \ref{summarytable}.
The hardness ratio and X-ray spectral analysis results show that the X-ray spectra of these LLAGN are consistent with power law models with photon indices ranging from 1--3, consistent with other AGNs \citep[e.g.,][]{dai04, saez08, she18, panagiotou20}.

\begin{deluxetable}{lccccc}
\label{summarytable}
\tablecaption{Summary of multi-wavelength properties of variability selected AGNs}
\tablecolumns{6}
\tablehead{\colhead{Name} & \colhead{W1$-$W2 (mag)} & \colhead{W3$-$W4 (mag)} & \colhead{$\log L_X$ (erg s$^{-1}$)} & \colhead{HR} & \colhead{$\Gamma$}}
\startdata
NGC 0863  & 0.417  & 2.723 & \nodata & \nodata & \nodata \\
UGC 02018 & -0.059 & 1.645 & \nodata & \nodata & \nodata \\
NGC 0988  & 0.029  & 1.782 & $<39.9^a$ & \nodata & \nodata \\
NGC 2552  & 0.112  & 3.883 & \nodata & \nodata & \nodata \\
UGC 05215 & -0.005 & 2.133 & $<41.3^a$ & \nodata & \nodata \\
NGC 3102  & -0.073 & 1.628 & \nodata & \nodata & \nodata \\
NGC 3499  & -0.050 & 1.260 & \nodata & \nodata & \nodata \\
NGC 3594  & -0.080 & 2.218 & \nodata & \nodata & \nodata \\
SDSS J112752.43+025038.3 & \nodata & \nodata & $<42.9^a$ & \nodata & \nodata \\
NGC 3835  & 0.026  & 2.135 & \nodata & \nodata & \nodata \\
NGC 3886  & -0.052 & 1.353 & $<41.3^a$ & \nodata & \nodata \\
NGC 3978  & 0.146  & 2.096 & \nodata & \nodata & \nodata \\
NGC 4041  & 0.269  & 2.651 & \nodata & \nodata & \nodata \\
NGC 4043  & 0.020  & 2.272 & $42.0^a$ & $-0.59\pm0.27^a$ & \nodata \\
NGC 4061  & -0.035 & 1.770 & $42.6^a$ & $-0.85\pm0.17^a$ & \nodata \\
NGC 4062  & -0.063 & 2.254 & $<39.7^a$ & \nodata & \nodata \\
NGC 4066  & -0.084 & 1.651 & $42.5^a$ & $-0.80\pm0.13^a$ & \nodata \\
NGC 4070  & -0.005 & 1.706 & $<41.5^a$ & \nodata & \nodata \\
NGC 4080  & -0.002 & 2.166 & $<39.1^a$ & \nodata & \nodata \\
NGC 4092  & 0.035  & 1.944 & $<41.2^a$ & \nodata & \nodata \\
NGC 4095  & -0.036 & 1.248 & $<41.4^a$ & \nodata & \nodata \\
NGC 4131  & -0.044 & 1.196 & $<40.8^a$ & \nodata & \nodata \\
NGC 4158  & 0.002  & 1.977 & $<40.6^a$ & \nodata & \nodata \\
NGC 4213  & -0.071 & 1.348 & $<41.3^a$ & \nodata & \nodata \\
NGC 4224  & 0.004  & 1.608 & $40.6^a$ & $-0.85\pm0.14^a$ & $2.3\pm0.4^b$ \\
NGC 4233  & -0.029 & 1.626 & $40.0^a$ & $-0.83^{+0.09}_{-0.06}$ $^b$ & $1.8\pm0.4^b$ \\
NGC 4235  & 0.340  & 2.055 & $42.6^a$ & $-0.75\pm0.01^a$ & \nodata \\
NGC 4241  & 0.018  & 2.183 & $<39.4^a$ & \nodata & \nodata \\
NGC 4272  & -0.025 & 1.884 & $<41.5^a$ & \nodata & \nodata \\
NGC 4944  & -0.056 & 1.398 & \nodata & \nodata & \nodata \\
NGC 5032  & -0.071 & 1.552 & \nodata & \nodata & \nodata \\
NGC 5162  & -0.085 & 2.276 & $41.9^a$ & $-0.49\pm0.28^a$ & \nodata \\
UGC 08516 & 0.036  & 2.357 & $<39.6^a$ & \nodata & \nodata \\
NGC 5273  & 0.172  & 2.510 & $40.7^b$  & $-0.96^{+0.18}_{-0.03}$ $^b$ & $3.1\pm1.5^b$ \\
IC 4345   & -0.069 & 1.430 & \nodata & \nodata & \nodata \\
NGC 5548  & 0.806  & 2.545 & \nodata & \nodata & \nodata \\
NGC 5633  & 0.117  & 2.143 & \nodata & \nodata & \nodata \\
\hline
\enddata
\tablecomments{$^a$Measured from eROSITA data. $^b$Measured from Chandra data. The X-ray energy range is 0.2--8 keV for eROSITA (0.2--5 keV for upper limits) and 0.2--10 keV for Chandra.}
\end{deluxetable}

\subsection{AGN environment}

When we check the WH24 catalog, we find that the number of all galaxies in the sample, variability selected LLAGNs, and luminous AGNs that are in a cluster are 620, 23, and 3, respectively, resulting LLAGN and luminous AGN fractions in the fields of  1.2\%/0.7\% and 3.7\%/0.5\% in the clusters. 
When we use the redshift range of $\Delta z = 0.03$ instead of $\Delta z = 0.05$, we find those included source numbers become 440, 21, and 3, respectively. While the number of AGNs in clusters is not strongly affected, the number of galaxies in clusters is reduced significantly, affecting the AGN fraction. Either result, however, displays significant difference between LLAGN fraction in clusters and fields.
We present results using the redshift range of $\Delta z = 0.05$ to minimize the potential incompleteness.

\begin{figure}
    \centering
    \includegraphics[width=6.5in]{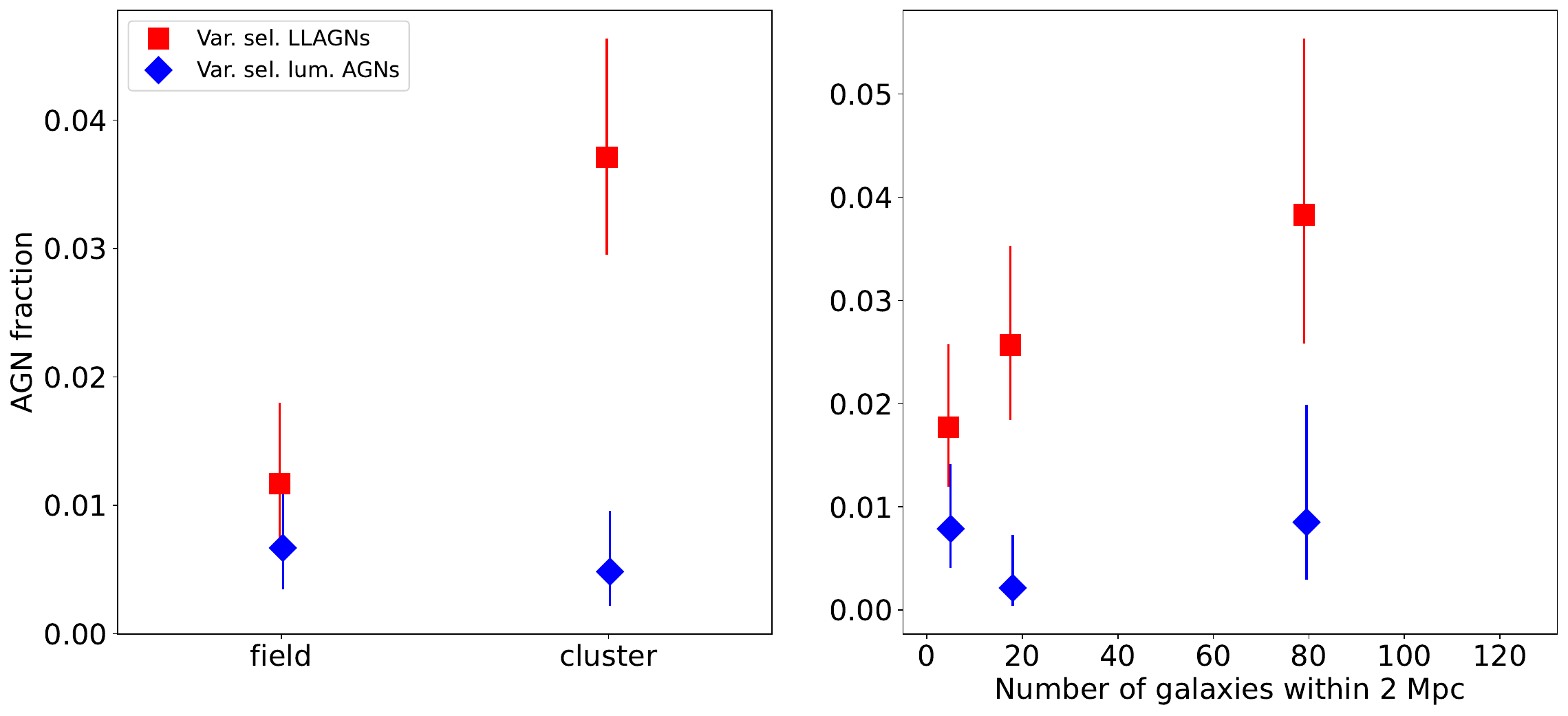}
    \caption{AGN fractions versus global environmental parameters of the association with WH24 galaxy clusters (left) and the number of NSA galaxies within 2 Mpc (right). The uncertainties are calculated by computing the confidence limit for binomial statistics for small number of events \citep{gehrels86}. The LLAGNs appear to be more prominent in denser environments, while luminous AGNs do not show such trend. The luminous AGN fraction may be underestimated for all bins due to the properties of the parent galaxy sample.}
    \label{globalenv}
\end{figure}

The number of NSA galaxies near variability selected AGN within 2~Mpc ranges from 2 to 110, with a median of 15 and a standard deviation of 21. We binned the number of nearby galaxies with bins that would maximize the number of AGNs in each bin. We observe a positive correlation between the LLAGN fraction and the number of galaxies in 2~Mpc. 
Both global environment parameters suggest that the LLAGN fraction is greater in clusters or denser environments, while such correlation is not observed for luminous AGNs (Figure \ref{globalenv}).

\begin{figure}
    \centering
    \includegraphics[width=6.5in]{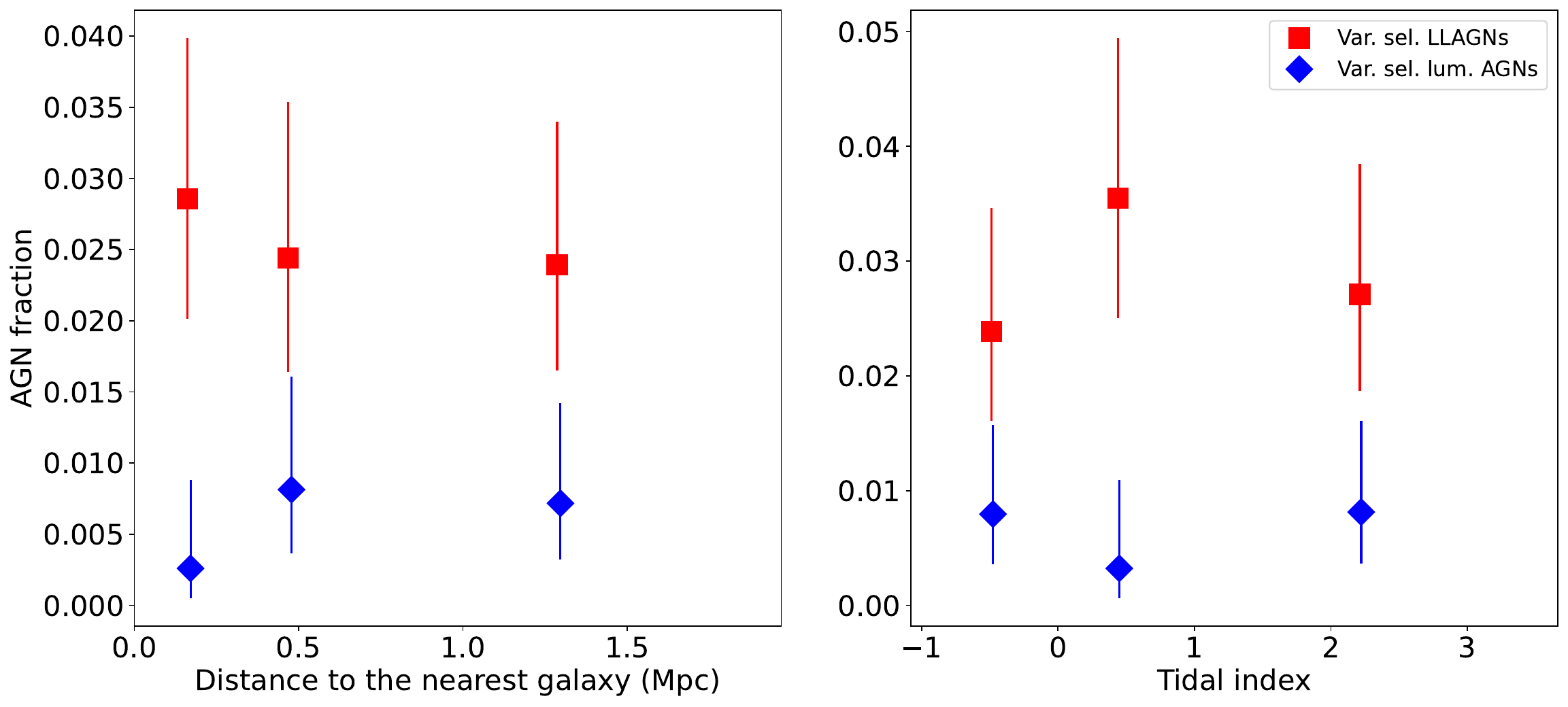}
    \caption{AGN fractions versus local environmental parameters of the distance to the nearest NSA galaxy (left) and the tidal index (right). The uncertainties are calculated by computing the confidence limit for binomial statistics for small number of events \citep{gehrels86}. Neither LLAGN fraction nor luminous AGN fraction displays a significant correlation with either parameter. The luminous AGN fraction may be underestimated for all bins due to the properties of the parent galaxy sample.}
    \label{localenv}
\end{figure}

The local environment parameters do not appear to have a significant correlation with the AGN fraction. The AGN fraction appears to be approximately constant with the distance to the nearest galaxy and the tidal index for both LLAGNs and luminous AGNs (Figure \ref{localenv}).

\section{Discussion and Conclusion}
\label{secdiscussion}

\begin{figure}
    \centering
    \includegraphics[width=4.5in]{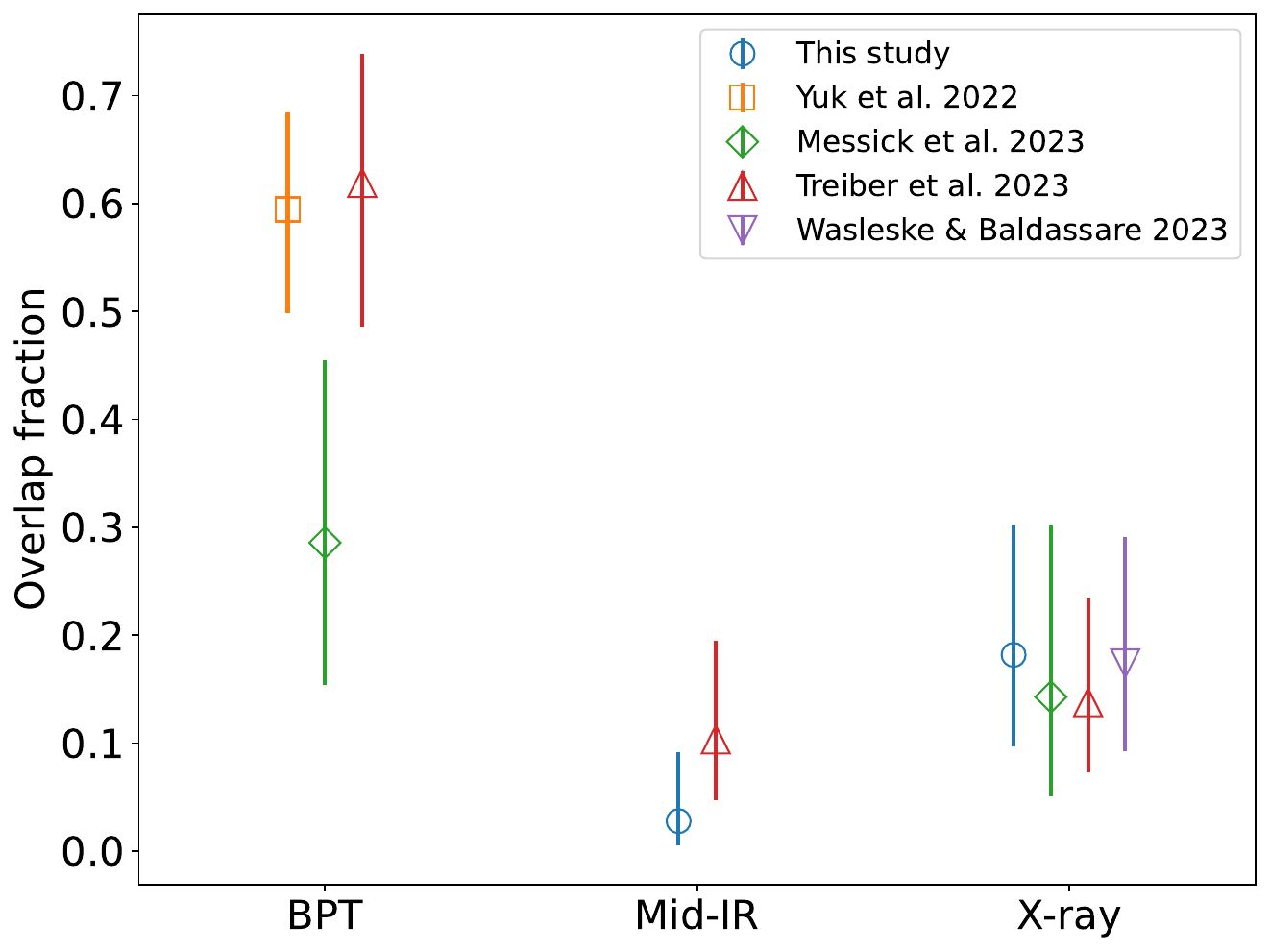}
    \caption{Summary of the fraction of variability selected AGNs that are classified as AGNs by other methods in the literature. The uniform standard of $L_X>10^{42}$ erg s$^{-1}$ is used for X-ray classifications. The uncertainties are calculated by computing the confidence limit for binomial statistics for small number of events \citep{gehrels86}. The overlap fraction between the variability selection method and other AGN selection methods ranges from a few percent with IR, 10--30\% X-rays, to 20--70\% optical line studies.}
    \label{overlap}
\end{figure}

From the multi-wavelength analysis of properties of LLAGN selected from the ASAS-SN survey, we find that the overlap between these AGNs from different selection methods is between a few to 60 percent. This is consistent with other analyses of variability selected LLAGNs.

\citet{wasleske23} analyzed the X-ray properties of 23 AGN candidates selected based on UV variability. 11 of their candidates had X-ray detection from Chandra and/or XMM-Newton. They compared the X-ray luminosity (0.5--7 keV) and the expected X-ray emission from star formation to determine whether the X-ray was coming from the AGN. They classified 10 of them to be X-ray AGNs.
\citet{messick23} performed a similar X-ray analysis on 14 nearby ($z<0.044$) low-mass ($M_*\lesssim5\times10^9 M_{\odot}$) variability selected AGNs. 4 of them had X-ray detections, all of which had X-ray luminosity (0.5--7 keV) greater than the anticipated emission from X-ray binaries. They also note that only 4 of their AGN candidates lie on AGN regime in the BPT diagrams.

\citet{treiber23} examined the variability selection method with other AGN selection methods. For their variability selected AGN sample, they analyzed Transiting Exoplanet Survey Satellite (TESS) light curves of 142,061 galaxies and identified 29 AGN candidates. They obtained optical spectra for 21 of them, and using BPT diagrams, they found 13 of them to be classified as AGNs. From mid-IR analysis, they found that 2 of their AGN candidates pass the \citet{assef10} criteria, 3 that pass the \citet{stern12} criterion. They also found that 4 of their candidates have the X-ray luminosity greater than $10^{42}$ erg s$^{-1}$.

The compilation of comparisons between variability and other AGN selection methods in the literature is shown in Figure \ref{overlap}. This series of multi-wavelength studies on variability selected AGNs supports the idea that the variability selection method is important for discovering LLAGNs that other methods may miss. 

\begin{figure}
    \centering
    \includegraphics[width=6.5in]{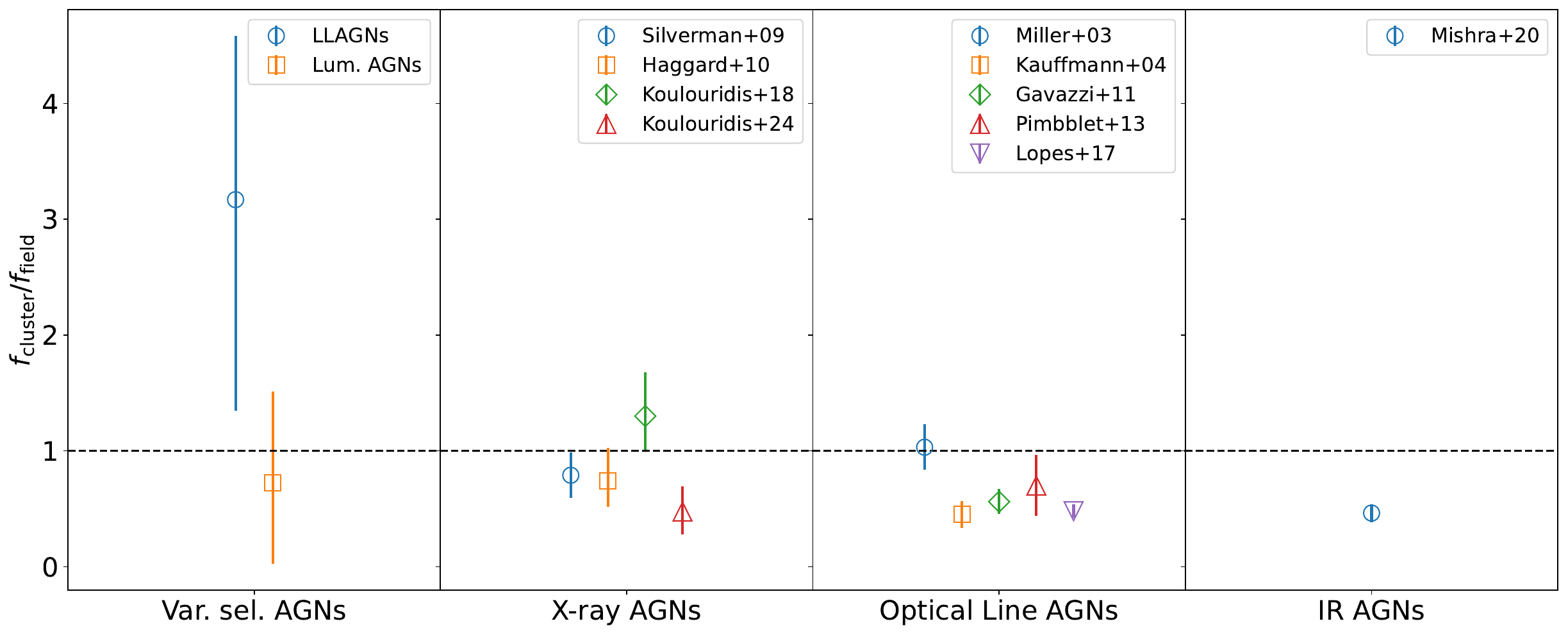}
    \caption{Compilation of cluster AGN fraction to field AGN fraction ratios from this work and past studies at low redshifts, classified by AGN selection methods.
    Previous studies using X-ray, optical lines, or IR AGN selection methods report that the cluster AGN fraction is in general lower than the field AGN fraction, which is consistent with our results with variability selected luminous AGNs. However, variability selected LLAGNs have much greater cluster AGN fraction than the field fraction.}
    \label{env_lit}
\end{figure}

We compare our environmental dependence of AGN fraction with previous studies at low redshift: \citealt{miller03}; \citealt{kauffmann04}; \citealt{silverman09}; \citealt{haggard10}; \citealt{gavazzi11}; \citealt{pimbblet13}; \citealt{lopes17}; \citealt{koulouridis18}; \citealt{mishra20}; \citealt{koulouridis24}. Specifically, the ratio of cluster AGN fraction to field AGN fraction. We note that not all of these works make clear cuts between clusters and fields. \citet{miller03}, \citet{kauffmann04}, \citet{silverman09}; and \citet{gavazzi11} compare regions with high and low local galaxy density. \citet{pimbblet13} and \citet{koulouridis24} use compare regions near and far from cluster centers. For comparing purposes, we categorize denser and less dense environments as clusters and fields, respectively. The comparison of AGN fraction ratios are shown in Figure \ref{env_lit}, where we further separate the environmental AGN fraction studies based on the AGN selection method.
Previous studies, mostly focusing on luminous AGNs, report that the AGN fraction in the field is greater than or comparable to the AGN fraction in clusters at low redshifts. Our result of variability selected luminous AGN fractions is consistent with previous studies. However, for variability selected LLAGNs, we found that the AGN fraction is higher in clusters.
It is possible that previous studies included many fewer, if at all, LLAGNs in their AGN sample, which may explain this discrepancy.  
The different environmental properties of LLAGNs may suggest different triggering mechanisms. For example, \citet{mishra20} speculate that fields allow more gas inflows during mergers without the ram pressure stripping effect in galaxy clusters, making field galaxies more likely to trigger AGNs at low redshifts. On the other hand, LLAGNs are thought to be more triggered by tidal disruptions and disk instabilities, which can be triggered by mergers \citep[e.g.,][]{hopkins09, hopkins14}.
This is consistent with our measurements of higher LLAGN fractions in the galaxy clusters, the global overdensity factor in a relatively large region; however, when only considering the local density factors, such as the tidal index or distance to the nearest galaxy, we did not measure significantly different AGN fractions on these local density parameters.
Our analysis suggests factors related to being inside global overdensity regions, e.g., velocity of galaxies before mergers and other factors, are especially important for contributing to the higher occurrence rate of LLAGNs in these regions.

\begin{acknowledgments}
H.Y. and X.D. would like to acknowledge NASA funds 80NSSC22K0488, 80NSSC23K0379 and NSF fund AAG2307802. H.Y. also thanks the Avenir fellowship.

This publication makes use of data products from the Wide-field Infrared Survey Explorer, which is a joint project of the University of California, Los Angeles, and the Jet Propulsion Laboratory/California Institute of Technology, funded by the National Aeronautics and Space Administration.

Parts of this work is based on data from eROSITA, the soft X-ray instrument aboard SRG, a joint Russian-German science mission supported by the Russian Space Agency (Roskosmos), in the interests of the Russian Academy of Sciences represented by its Space Research Institute (IKI), and the Deutsches Zentrum für Luft- und Raumfahrt (DLR). The SRG spacecraft was built by Lavochkin Association (NPOL) and its subcontractors, and is operated by NPOL with support from the Max Planck Institute for Extraterrestrial Physics (MPE). The development and construction of the eROSITA X-ray instrument was led by MPE, with contributions from the Dr. Karl Remeis Observatory Bamberg \& ECAP (FAU Erlangen-Nuernberg), the University of Hamburg Observatory, the Leibniz Institute for Astrophysics Potsdam (AIP), and the Institute for Astronomy and Astrophysics of the University of Tübingen, with the support of DLR and the Max Planck Society. The Argelander Institute for Astronomy of the University of Bonn and the Ludwig Maximilians Universität Munich also participated in the science preparation for eROSITA.

\end{acknowledgments}

\bibliographystyle{apj}
\bibliography{reference}

\end{document}